# Insights on the variability of Cu filament formation in the SiO$_2$ electrolyte of quantized-conductance conductive bridge random access memory devices

Florian Maudet[1, *], Veeresh Deshpande[1], Catherine Dubourdieu[1, 2,*]

1. Institute Functional Oxides for Energy-Efficient Information Technology, Helmholtz-Zentrum Berlin für Materialien und Energie, Hahn-Meitner Platz 1, 14109 Berlin, Germany

2. Freie Universität Berlin, Physical Chemistry, Arnimallee 22, 14195 Berlin Germany

Corresponding author: catherine.dubourdieu@helmholtz-berlin.de, florian.maudet@helmholtz-berlin.de

**Abstract**

Conductive bridge random access memory devices such as Cu/SiO$_2$/W are promising candidates for applications in neuromorphic computing due to their fast, low-voltage switching, multiple-conductance states, scalability, low off-current, and full compatibility with advanced Si CMOS technologies. The conductance states, which can be quantized, originate from the formation of a Cu filament in the SiO$_2$ electrolyte due to cation-migration-based electrochemical processes. A major challenge related to the filamentary nature is the strong variability of the voltage required to switch the device to its conducting state. Here, based on a statistical analysis of more than hundred fifty Cu/SiO$_2$/W devices, we point to the key role of the activation energy distribution for copper ion diffusion in the amorphous SiO$_2$. The cycle-to-cycle variability is modeled well when considering the theoretical energy landscape for Cu diffusion paths to grow the filament. Perspectives of this work point to developing strategies to narrow the distribution of activation energies in amorphous SiO$_2$.

# Introduction

Recently extensive work has been conducted to study and enhance the performances of resistive switching memristive devices as they are promising candidates for the next generation of nonvolatile random access memories or for neuromorphic applications [1–3]. Of all the studied configurations, conductive bridge random access memory (CBRAM), also named electrochemical metallization cell (ECM) holds great potential as it offers the possibility of fast switching, multiple states, ultimate scaling for ultra large scale integration, and low power consumption [4–6]. The possibility in such systems to obtain multiple state resistance values is of great interest for neuromorphic applications. $SiO_2$-based CBRAM have been studied with a variety of active electrodes, mainly Ag and Cu [7–10] and more recently Co [11]. The different resistance states are governed by the formation or dissolution of a metallic filament (Cu, Ag…) in the dielectric $SiO_2$ solid electrolyte [7,8,12]. When a positive bias is applied (on the active electrode), metallic ions migrate through the solid electrolyte which leads to the electrodeposition of the metal (Cu, Ag…) on the passive W electrode. Once the conductive filament bridges both electrodes, the device is in a low resistive state. Upon a negative bias, the metallic filament is dissolved by the migration of metallic ions (Cu, Ag…) out of the filament [13]. The $Cu/SiO_2/W$ system offers the advantage of being fully CMOS compatible [14–22]. This stack was shown to exhibit quantized quantum conductance states and a remarkably low operating power [21]. One of the major challenge for the integration of CBRAM in practical applications remains their stochasticity [4]. There is a lack of systematic statistical studies in literature for a clear understanding of these devices.

In this paper, we first present an analysis of the quantum conductance state distribution in $Cu/SiO_2/W$ CBRAM devices and show that it varies quite significantly when increasing the programming current. To understand the origin of this variability, we then focus on the I-V curves and particularly on the SET voltage, which relates to the Cu filament formation. The cycle-to-cycle SET voltage variability is measured on more than hundred fifty devices and analyzed with a physical model.

**Results**

The stack in our devices consist of a 100 nm W bottom inert electrode deposited by sputtering on a p-type Si substrate with 300 nm thermal oxide, a 10 nm $SiO_2$ film deposited by plasma-enhanced chemical vapor deposition as the electrolyte and a top 50 nm Cu active electrode deposited by sputtering (Fig. 1(a)). The top electrode is capped with 100 nm Au (with a 10 nm Ti seed layer underneath) to avoid oxidation of the Cu electrode with time. The active $Cu/SiO_2/W$ device is a cross-point of 30 µm x 30 µm, fabricated using conventional photolithography with unpatterned bottom electrode (Fig. 1(b)).

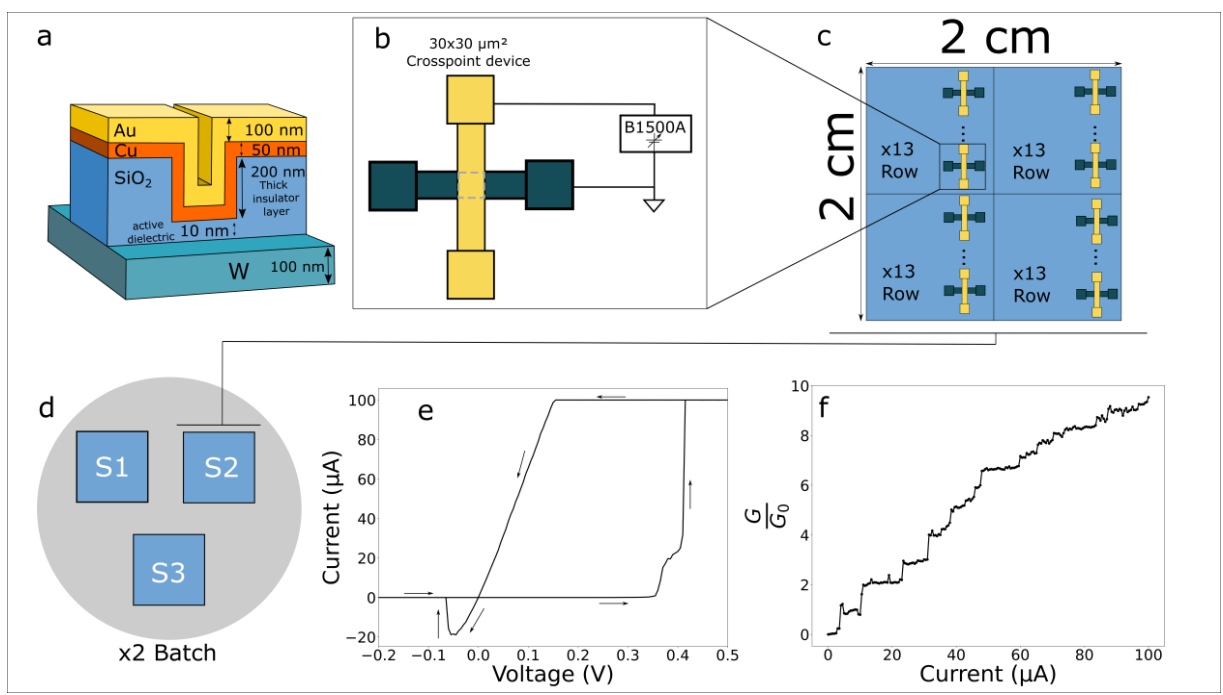

*Figure 1 Devices under study. (a) Schematic of the device stack. (b) Cross point top view with measurement scheme. (c) Typical 2x2 cm$^2$ Si samples on which the devices are prepared with 4 dies and 13 devices on each die. (d) Positioning of the samples in each batch preparation. (e) Typical pinched hysteresis I-V curve measured with a compliance current of 100µA and plotted in linear scale. (f) Example of quantum conductance curve versus current observed in the devices.*

For the statistical analysis, a batch consisting of three samples was processed, each sample being composed of 4 dies and each die of 13 devices (Fig. 1 (b) and (d)), hence providing a total of 156 devices. The samples were stored under $N_2$ atmosphere with desiccant. A pinched hysteresis I-V curve characteristic of the bipolar resistive switching in our CBRAMs is shown in Fig. 1(c) for a typical device. During a positive sweep, the devices transition (SET transition) from a high resistive state (HRS) – the OFF state - to a low resistive state (LRS) – the ON state. No forming step is required to form the first filament. Very low voltages, typically below 450 mV are required for the SET operation (compliance current of 100 µA). Upon negative voltage sweep, the devices are RESET (transition from LRS to HRS) at voltages typically around -100 mV. As previously reported [21], these devices exhibit half-integer quantized quantum conductance states that are clearly evidenced when performing a current sweep as shown in Fig. 1 (f). As a matter of fact, not only half-integer but also quarter-integer quantum conductance states are observed, whose origin will be discussed elsewhere.

*Variability of the quantum conductance states*

The occurrence of several well-defined conductance states is of high interest for neuromorphic computation. However, for practical use, the states should be distinguishable and controllable. We show in Fig. 2 the distribution of the quantum conductance values measured for different programming currents ranging from 10 to 100 µA (with a compliance voltage of 1V) on a total of 26 devices (2 dies) with 30 cycles performed on each device.

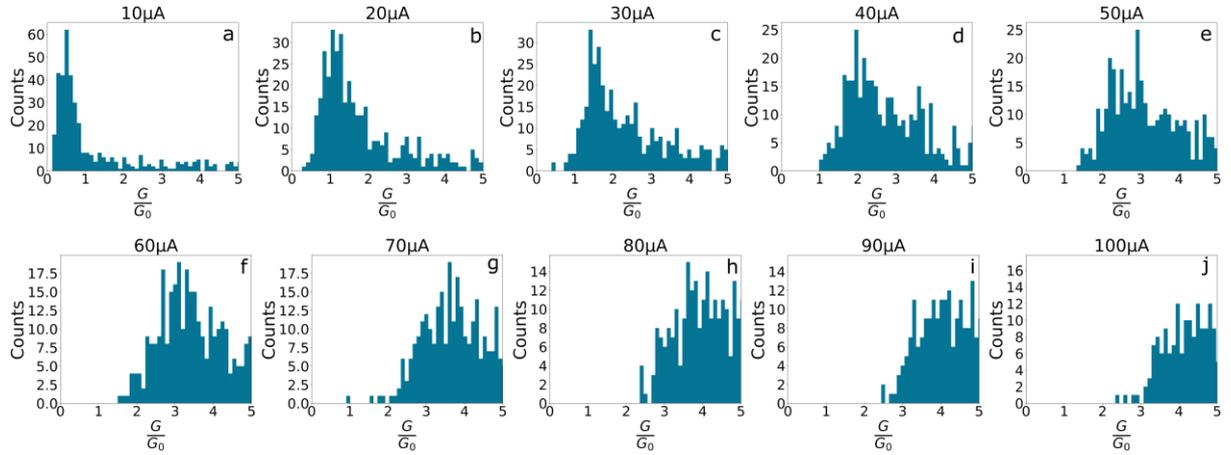

*Figure 2* Quantum conductance state distribution for compliance current of : (a) 10μA , (b) 20μA, (c) 30μA, (d) 40μA, (e) 50μA, (f) 60μA, (g) 70μA, (h) 80μA, (i) 90μA and (j) 100μA. The distribution is cumulative of all devices and all cycles performed.

For a low programming current of 10 μA (Fig. 2(a)) the quantum conductance state distribution is relatively narrow and centered at 0.5 $G_0$, suggesting that this state can repeatably be programmed using a 10 μA current. For an increasing programming current, the distribution shifts to higher conductance states, and gets broader. The quantum conductance G is related to the diameter of the filament formed [21]. G values are clearly increased for larger currents but also appear more random. At least three differentiable memory states, (0.5-1$G_0$, 1.5-3$G_0$ and 3-5$G_0$) are available in the programming current range of 10-100 μA. The broadening of the quantum conductance state distribution clearly limits the potential of such devices. To improve this aspect, it is therefore key to understand the origin of the variability in such devices.

*Variability of SET voltages*

We have focused our analysis on the SET voltages of the devices, which correspond to the transition from OFF to ON states. The device-to-device, die-to-die or sample-to-sample variations have been studied on a large number of devices (156 devices) in order to distinguish the intrinsic variability - clearly observed upon cycle-to-cycle measurements - from potential extrinsic variability. Cycle-to-cycle variation is associated with the intrinsic physical stochasticity of the device. Device-to-device/die-to-die and especially sample-to-sample variations originate not only from the intrinsic stochasticity but also from fabrication process related effects like thickness inhomogeneity, deposition processes variations, processing residues, etc. For each of these devices, thirty full I-V cycles were measured and for the statistical analysis, only devices showing at least fifteen full I-V cycles were considered.

In Fig. 3(a) thirty consecutive I-V sweep cycles are shown for a representative device. The device exhibits a high $R_{ON}/R_{OFF}$ ratio of ~ $10^6$ obtained repeatedly and a relatively narrow SET and RESET voltage distribution. This repeatability implies that the RESET process is complete, *i.e.* the conductive paths are completely removed in the RESET sweep or are below the detection limit. An incomplete RESET would indeed lead to a lowering of the SET voltage in the succeeding SET sweep. In Fig. 3(b), the corresponding SET and RESET voltages are presented. Low median values of the SET voltage (0.385±0.062 V) and RESET voltage (-0.081±0.032 V) are obtained. For all measured devices, the first cycle is not different from the following ones and stays within the overall voltage distribution. This observation supports the completeness of the RESET process. Now, we consider the SET and RESET distribution of thirty cycles on each device for different devices of a single die, of different dies, of different samples as shown in Fig. 3(c), (d), (e and f), respectively. In Fig. 3(c), the boxplot underlines the relatively high homogeneity of the devices on a single die, with the exception of one or two of them. The four dies (Fig. 3(d)) of one sample exhibit consistently reproducible values of voltage median and distribution in SET and RESET voltages, apart from one die (D43) that has a slightly broader distribution of the SET voltage. This observation highlights the importance of the analysis of a large number of devices as the sole analysis of the die D43 could have led to a skewed picture. A relatively good sample-to-sample repeatability is also observed for the three samples processed within the same batch (Fig. 3 (e), (f)).

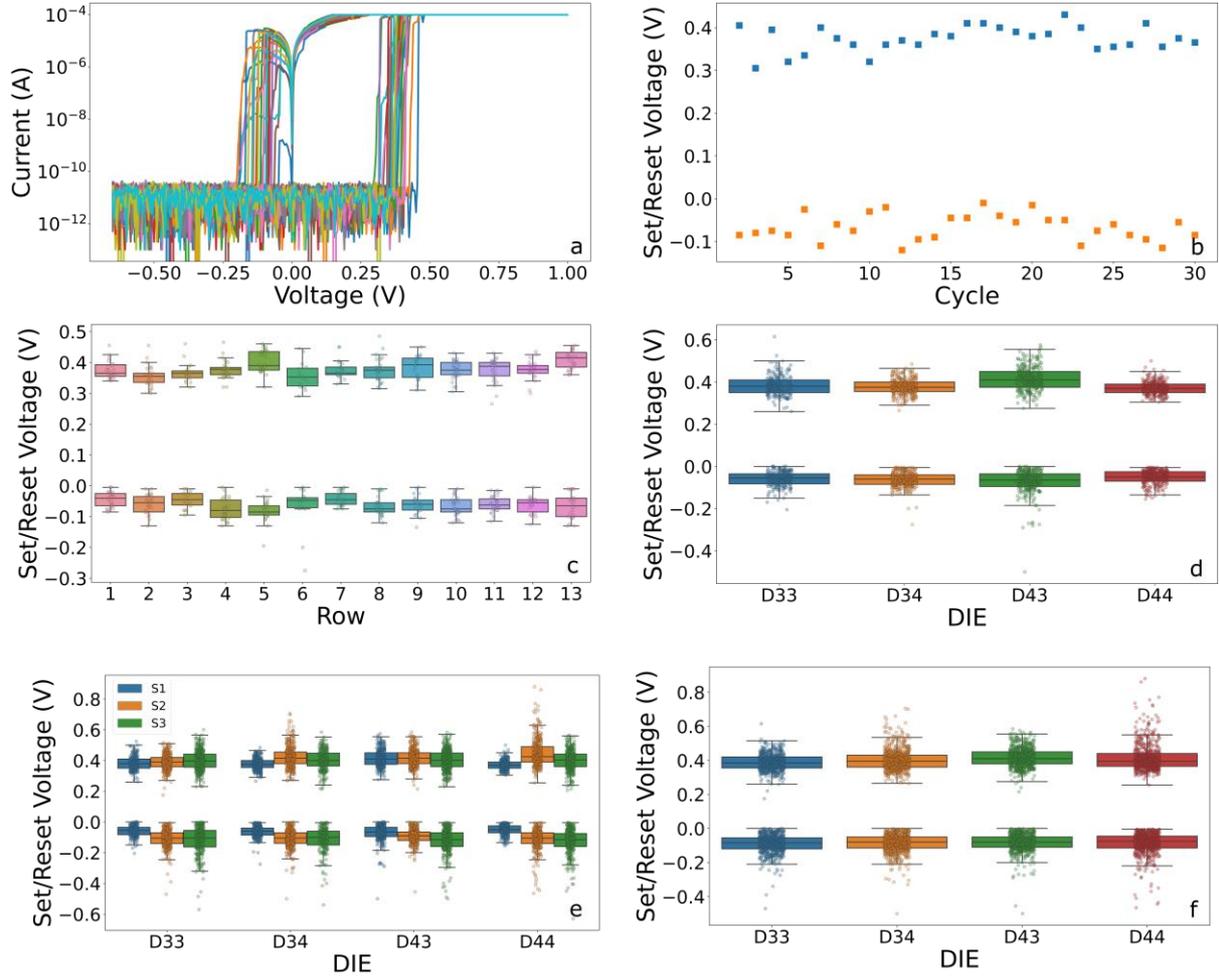

*Figure 3. Statistical analysis of I-V sweep measurements. (a) Thirty I-V cycles of a representative device. (b) SET and RESET voltages of the device shown in (a). (c) SET and RESET voltages obtained for 13 devices from one die. (d) SET and RESET voltages for 4 dies of a same sample – (e) SET and RESET voltages for 4 dies of three samples of the same batch (total of 156 devices) – (f) SET and RESET voltages for all 156 devices. For (c)-(f), the rectangles of the boxplots represent the first and third quartiles and the bars the 5th and 95th centiles.*

To summarize the results on the switching variability - from OFF to ON states – of the 156 devices under study, we give in Table 1 the median value and the standard deviation of their SET voltage values ($\sigma_{V_{set}}$) and the difference between the third and first quartile ($Q_3$-$Q_1$) for different sampling size, representative of the cycle-to-cycle, device-to-device, die-to-die and sample-to-sample variabilities.

|  | 30 cycles<br>1 device | 1 die<br>13 devices | 4 dies = 1 sample<br>52 devices | Batch = 3 samples<br>156 devices |
|---|---|---|---|---|
| $Median_{V_{set}}$ | 0.385 V | 0.410 V | 0.410 V | 0.395 V |
| $\sigma_{V_{set}}$ | 0.062 V | 0.065 V | 0.069 V | 0.065 V |
| $Q_3-Q_1$ | 0.071 V | 0.075 V | 0.075 V | 0.070 V |

*Table 1 : Median value, standard deviation and difference between the third and first quartile of the SET voltage values for different sampling size. For each device, 30 cycles were measured (a batch of 3 samples represents 4680 measurements).*

The median SET value of the different devices is comprised between 385 and 410 mV with a low dependence on the sampling size indicating a low device-to-device variability. A relatively low cycle-to-cycle variation of the SET voltages is observed for a single device (0.062 V). The standard deviation observed for one die (0.065 V) is very close to the dispersion of a single device. For the four dies of a sample as well as for the three samples in the batch, similar values of SET voltage dispersion $\sigma_{V_{set}}$ are observed as compared to the device-to-device one on a single die, with 0.069 V and 0.065 V respectively. The similar trend observed for the difference between the first and third quartile is indicative of a normal distribution of the set values. These observations confirm that the cycle-to-cycle variability is the major cause of the dispersion of the SET values and that the fabrication can be controlled to achieve homogenous samples leading to a negligible device-to-device variation.

**Discussion on the SET voltage distribution**

It is now well admitted that switching of these $Cu/SiO_2/W$ devices is governed by the formation of a Cu metallic filament as a result of oxido-reduction processes occurring at the active and passive electrodes [23,24]. Once the copper ions are formed under electrochemical oxidation, they diffuse through the electrolyte $SiO_2$ under the applied electric field [25]. For this to happen copper ions must have enough energy to overcome the energy barriers of the different sites encountered in the amorphous $SiO_2$ electrolyte where the local environment varies (Fig 4(a)), which inherently leads to stochasticity. Guzman *et al.* have modelled the migration of copper atoms and clusters in amorphous $SiO_2$ using density functional theory and report a broad distribution of activation energies for Cu migration, ranging from 0.2 to 1.2 eV as shown in Fig 4(b) [26]. Hence, for different nucleation position of the filament, copper ion migration will happen at different rates.

A model can be developed to analyze the influence of this distribution of activation energies on the SET voltage distribution. Different approaches have been proposed to model the behavior of CBRAM devices: finite elements, kinetic Monte Carlo, analytical, and compact models [10,27,28]. Among them, analytical models constitute a relatively simple approach to predict the impact of physical properties of the material on the device performances. When taking into account all physical mechanisms at play for the device behavior, analytical models have to be solved numerically as it leads to equations with an implicit form [27]. However, under reasonable assumptions these equations can be solved analytically. Here, to find out the impact of the distribution of activation energies on the distribution of SET voltages, an analytical model assuming that migration of Cu is the limiting factor for the filament formation was used [28]. Five hypotheses are made, defining the range of validity of this model. First, we consider that the devices do not have a cycle history *i.e.* the RESET is complete, which is supported by the repeatability of the OFF state (RESET voltage) and by the absence of forming voltage, suggesting the formation of a new filament for each cycle. Second, the influence of temperature on the filament is neglected as we consider only SET voltages happening under low current (< nA) and thus low joule heating effect. Third, copper migration is supposed to be the limiting mechanism of the switching process. Fourth, the filament is assume to grow vertically before lateral expansion occurs [29]. Finally, the nucleation of the filament is assumed to happen at random locations as it will be sensitive to noise in the electric field.

Following these assumptions the growth rate of a cylindrical filament of height $h$ can be modeled by the following Arrhenius law [28,30]:

$$\frac{dh(t)}{dt} = A\, e^{\frac{-(E_a - q\alpha V)}{kT}} = A\, e^{\frac{-(E_a - q\alpha\beta t)}{kT}} \quad (1)$$

where the constant A is an exponential prefactor, $E_a$ the activation energy for Cu ion migration (eV), V is the voltage drop across the gap between the filament and the inert electrode, q is the electron charge, α is the barrier lowering factor and T is the temperature of the ions. As the applied voltage is swept, the factor $\beta t$ is introduced to take into account the voltage dependence over time where $\beta$ is the voltage sweep rate (V.s$^{-1}$), t is the time (s).

The height $h$ is then directly expressed by:

$$h(t) = \frac{v_h kT}{q\alpha\beta}\, e^{\frac{-(E_a - \alpha q\beta t)}{kT}} \quad (2)$$

We define the SET point of the device for $h = L$ where L is the SiO$_2$ dielectric thickness, *i.e.* when the filament has bridged top and bottom electrodes. Consequently, the SET voltage is expressed by:

$$V_{set} = \frac{E_a + \ln\left(\frac{Lq\alpha\beta}{v_h kT}\right) kT}{q\alpha} \quad (3)$$

*This expression allows us to compute the SET voltage distribution in our devices. The parameter used for the calculation of V$_{set}$ are summarized in*

**Table 2.** . The length $L$ corresponds to the thickness of the film and $\beta$ is the sweep rate of our measurement. A schematic of the model is presented in Fig. 4(c). As for $E_a$, we have considered the distribution of activation energies calculated by Guzman *et al.* for copper ion diffusion in amorphous SiO$_2$ [26]. We fitted the distribution reported in ref. 27 with two gaussians (G1 and G2) as shown on Fig. 4(b). Only $A$ and $\alpha$ are free parameters, which were adjusted so that the median SET value of the model corresponds to the one of the full set of devices. We found $A = 5.10^{-7}$ m.s$^{-1}$ which is similar to the value of $8.10^{-7}$ m.s$^{-1}$ that is calculated from ref. 20 assuming a diffusion coefficient of Cu in SiO$_2$ of $D_0 = 1.10^{-10}$ cm$^2$.s$^{-1}$. To our knowledge no value is available for α for comparison but the obtained value of 0.95 is physical ($0 < α < 1$) and reasonable with the thickness considered.

Using equation (3) the variance of the SET voltage can be determined analytically. In the ideal case of a perfectly smooth surface ($\sigma_L = 0$) and homogenous sample ($\sigma_{v_h} = 0$) it is expressed as:

$$\sigma_{V_{set}} = \frac{\sigma_{E_a}}{q\alpha} \quad (4)$$

This constitutes the minimum intrinsic variability of the SET voltage of a device in this model.

| Symbol | Value |
| --- | --- |
| $E_a$ | $G1(A = 1.8, \mu = 0.43\ eV, \sigma = 0.06\ eV)$ <br> $G2(A = 1.7, \mu = 0.73\ eV, \sigma = 0.16 eV)$ |
| $A$ | $5.10^{-7}$ m.s$^{-1}$ |
| $\alpha$ | 0.95 |
| $\beta$ | $2.5.10^{-2}$ V.s$^{-1}$ |
| $L$ | 10 nm |

*Table 2. Values of the parameters used for the analytical model*

In the calculation of the SET voltage cumulative distributions shown in Fig. 4(d) and (g), we have considered two cases: the distribution of the activation energy $E_a$ is represented by the sum of the two gaussian functions G1+G2 or by only the G1 gaussian. The G1+G2 gaussian of activation energies lead of course to a much larger SET voltage variability (80% of the SET values are comprised between 0.34 and 0.87 V) as compared to the one obtained with G1 alone (0.34 and 0.48 V). On Fig. 4(e) and (h) we represent the experimental cumulative distributions of the SET voltages for thirteen devices (one die) and a calculated curve using G1 and G1+G2 respectively. Most of the measured cumulative SET distributions are very well modelled using only the G1 function. This result indicates that at least one diffusion path with an energy landscape that does not exceed 0.43 eV is always available for the Cu ions diffusing in the $SiO_2$ electrolyte so that this path is favored. A very good match is also obtained for the concatenated results of the cumulative distribution functions of a die and of the 156 devices (Fig. 4 (i), (f)). The simulation based on the full gaussian distribution of the activation energy (G1+G2) only represents few cases as shown in Fig. 4(h). Of the total of 156 devices, 9 devices exhibited a SET value larger than 0.60 V.

The theoretical standard deviation for the G1 distribution is calculated to be $\sigma_{V_{set\,G1}} = 0.063$ V close to the standard deviation observed experimentally for different sampling size (single device, single die, single sample, and full batch) as shown in Table 1. The theoretical standard deviation of G1+G2 is calculated to be $\sigma_{V_{set\,G1+G2}} = 0.175$ V representing a larger distribution of the SET voltage.

The good agreement between experiments and the calculated $V_{set}$ distributions indicates that the variability originates – at least to a large extent - from the distribution of activation energies centered about 0.43 eV for the diffusion of copper ions in the $SiO_2$ amorphous matrix.

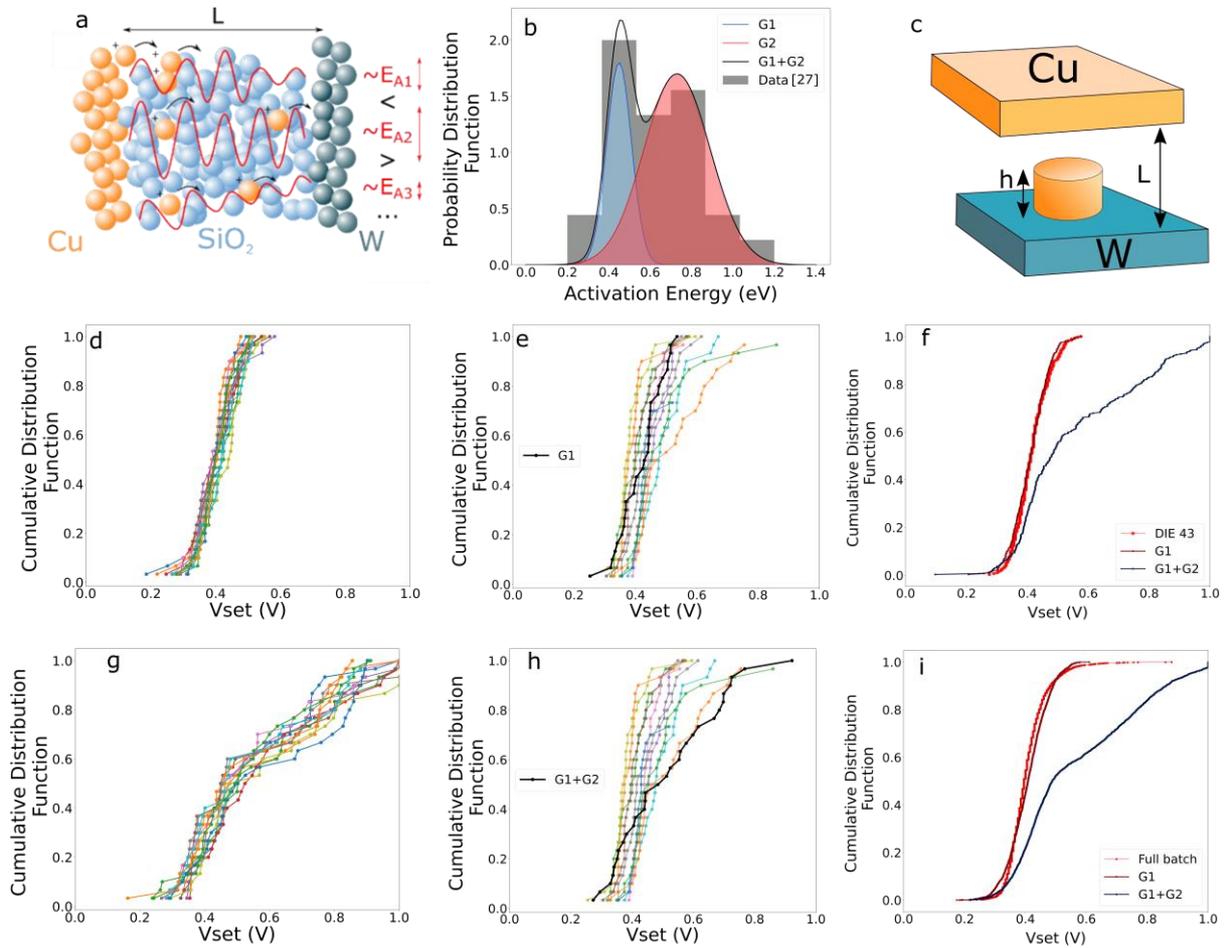

*Figure 4:* Modelling of the variability of the SET voltages (a) Schematic representation of the Cu ion drift process. (b) Probability distribution function calculated from [26] with the associated fit of the activation energy distribution with two Gaussians. The fitted distribution is used in our analytical model. (c) Schematic depicting the analytical model used for filament growth. Simulated cumulated distribution function of 13 devices (30 cycles) considering the gaussian distribution G1 (d) or G1+G2 (e). Cumulative distribution function of the SET voltage for 30 cycles of the 13 devices of a die (f,g), of a single die (h) and of the three samples (i). For each of the measurement, a simulation example is made for the same number of cycles considering only G1 (f, h and i) and G1+G2 (g, h and i). For the activation energy (see Fig. 4(b)).

**Conclusion**

We characterized and analyzed the stochasticity of the SET voltages of 156 Cu/SiO$_2$/W CBRAM devices. For the electrolyte produced here by PECVD at low temperature (150°C), the median SET value of the 156 devices is of 0.395 V with a standard deviation of 0.065 V, whereas the cycle-to-cycle standard deviation is of 0.062 V. The cycle-to-cycle variability was examined using an analytical model including a distribution of activation energies for Cu ion diffusion in amorphous SiO$_2$, from a reported DFT calculation. The cycle-to-cycle stochastic SET behavior appears intrinsically related to the energy landscape of the diffusion paths available in the amorphous SiO$_2$ and is reproduced very well by considering diffusion paths with a relatively low variability in energy (0.060V) and an average barrier height of 0.43 eV. Only few devices (9 out of 156) exhibit a behavior indicative of a much larger activation energy distribution. The minimization of the available diffusion paths in SiO$_2$ is key to narrow the variability of the SET and RESET voltages. With this statistical study, we also show that the device-to-device variability is similar to the cycle-to-cycle one, which is the result of a controlled process of the device fabrication. Future perspective of this work includes the integration of such devices in the back-end-of-line of CMOS chips and on flexible substrates for the development of stochastic neuromorphic systems [31].

## Methods

**Device fabrication.** A 100 nm W layer was deposited by sputtering at room temperature on 300 nm thermally-grown SiO2 on Si substrates. A thick insulator layer of 200 nm of SiO2 was then deposited by PECVD at 120 °C and patterned by lift-off using photolithography. A blanket deposition of 10 nm SiO2 was performed also by PECVD at 120 °C for the active electrolyte layer. Then, 50 nm Cu followed by 5 nm Ti / 100 nm Au capping layers were deposited by thermal evaporation and patterned by lift-off to create 30x30 µm crosspoint. Finally, the bottom pads were opened to guarantee a good ohmic contact to the bottom electrode by reactive ion etching of the 10 nm SiO2 layer using CHF3/Ar.

**Electrical measurements.** Electrical measurements were performed on a MPI TS2000-SE probe station with Keysight B1500A semiconductor parameter analyzer on 30x30 µm² cross point devices. Voltage-current measurements (30 cycles for each device) were performed for the determination of the quantized conductance on the thirteen devices for two dies on one sample. The current was swept from 10 µA to 100 µA with a compliance voltage of 1 V. For the SET value study, current-voltage measurement were performed (30 cycles for each device) with current compliance of 100 µA and a sweep rate of 0.025 V.s-1 on each of the thirteen devices of the four dies for three samples.

**Acknowledgment**

All authors acknowledge Stefan Bock, Luca Sulmoni and Ronny Schmidt for technical support at the clean room of the Physics Department of the Technische Universität Berlin.

**Author contributions**

C.D. and V.D. conceived the study. F.M. and V.D. prepared the devices and performed the electrical measurements. F.M. performed the data analysis. All authors participated in the discussion of the results. F.M. and C.D. wrote the manuscript and V.D. edited the manuscript.

**Data availability**

The data that support the findings of this study are available from the corresponding authors upon reasonable request.

**Competing Interests**

The authors declare no competing interests.

**Additional information**